# Interfilament Resistance at 77 K in Striated HTS Coated Conductors

R. Gyuráki, A. Godfrin, A. Jung, A. Kario, R. Nast, E. Demenčik, W. Goldacker, F. Grilli

*Abstract* - Striating HTS coated conductor (CC) tapes into narrow filaments offers the possibility of reducing the tapes' magnetization losses without unreasonably decreasing their current-carrying capability. However, realizing well-separated striations presents technological challenges, especially if the number of filaments is large and/or if a thick layer of metallic stabilizer is present. In these situations, the filaments can be easily coupled and their effectiveness to reduce magnetization losses is strongly diminished or eliminated. While the onset of coupling is well visible from magnetization loss measurements, the actual path of the coupling current is unknown. In this contribution we present a systematic study of the transverse resistance in HTS CC samples in order to get a deeper understanding of those paths. The measured samples differ in terms of manufacturer (SuperPower and SuperOx), and presence and thickness of stabilizer material. In addition, oxidation is used as a means to increase the resistance between the filaments in non-stabilized samples. The results are interpreted with a chain network model. This work provides useful insights on the factors determining the transverse resistance in striated HTS CCs, thus indicating ways to improve the effectiveness of the striation process for AC loss reduction.

## I. INTRODUCTION

HIGH Temperature Superconductors (HTS), such as REBCO coated conductors are in the focus of active research. This is due to their excellent critical current performance at liquid nitrogen temperatures, which make them an ideal candidate for applications such as superconducting motors [1] and generators [2]. These type of applications are driven with AC current which further result in AC loss in REBCO conductors. In recent years these conductors are available in long length [3], [4], which are suitable for applications and are offered by several companies, suchs as SuperPower [3], SuperOx [5], Fujikura [4], SuNAM [6] and American Superconductor [7].

The authors are with the Institute of Technical Physics, Karlsruhe Institute of Technology, 76344 Eggenstein-Leopoldshafen, Germany (e-mail: roland.gyuraki@kit.edu; aurelien.godfrin@kit.edu; alexandra.jung@kit.edu; anna.kario@kit.edu; rainer.nast@kit.edu; eduard.demencik@kit.edu; wilfried.Goldacker@kit.edu; francesco.grilli@kit.edu).

It has been long proved that striating an HTS tape into electrically insulated filaments by means of a laser scribing reduces the magnetization losses (see for example [8]). The theoretical loss reduction is directly proportional to the number of filaments. However, depending on the scribing parameters and frequency of the applied field, the measured AC losses in striated tapes can be larger than expected from the filamentary structure. This is mostly due to the coupling of filaments caused by conductive material bridges between neighbouring strips [9]. It is therefore important to control the electrical resistance between the filaments to balance coupling effects and current sharing.

Transverse resistance is a key parameter that influences the coupling loss [10]–[12] and furthermore allows determining the current paths between the filaments of HTS tapes. Oberly *et al.* [2] and Carr [13] have expressed the coupling loss per unit volume in terms of the transverse resistivity as,

$$\frac{\overline{P_c}}{V_b} = \frac{1}{4\rho_{tr}} \left( f\, B_p\, L \right)^2 \tag{1}$$

Where $\overline{P_c}$ is the r.m.s. power of the coupling loss, $V_b$ is the barrier volume, $f$ is the frequency, $B_p$ is the peak magnetic field intensity and $L$ is the twist pitch. $\rho_{tr}$ is the transverse resistivity defined as,

$$\rho_{tr} = \rho_b \left[ \frac{(n-1)W_b}{(n-1)W_b + nW_{YBCO}} \right] \tag{2}$$

Where $\rho_b$ is the barrier resistivity, n the number of filaments and $W_b$ and $W_{YBCO}$ are the barrier and YBCO widths, respectively.

According to Oberly *et al.* [14] the interfilamentary barrier resistance is the most important factor for reducing the ac loss in applications such as motors and generators. Despite this there have not been many works investigating this topic in detail [10], [14]–[18].

In this paper the transverse resistance between filaments of SuperOx and SuperPower tapes, with and without copper stabilization has been investigated. The tapes were prepared with a more optimized scribing method, compared to previous work from our institute [10][15]. The experiments have also been extended to a wider range of Ag- and Cu-cap tapes with different preparation methods. The measurement data was used to describe possible current paths and to infer coupling effects, important for ac losses calculation [18]. Finally, the merits of a simple equivalent circuit transverse resistance model are shown.

## II. Sample Preparation

Striated samples using 12 mm wide and 15 cm long tapes, based on Hastelloy substrate, provided by SuperPower and SuperOx company were prepared. 10 filaments were done on the samples using laser micromachining with a groove width of 20 μm following the procedure described by Nast *et al.* [19].

Most striations were prepared using Cu stabilised tapes with optimised laser parameters for the given copper thickness. Samples prepared with this method will be referred to as "Striated After Electroplating" (SAE). The second method used was the striation of the Ag-cap tapes, followed by oxidation and electroplating of copper (in the case of SuperOx samples). Samples prepared with this second method will be referred to as "Striated Before Electroplating" (SBE). Oxidation was done in pure oxygen atmosphere at 500 °C for one hour to oxidize the laser grooves (Hastelloy) and thereby increase the resistivity of the groove [19]–[21].

## III. Methods

Measurements of transverse resistance were done in two different configurations: filament-to-filament resistance ($R_{FTF}$) and profile resistance ($R_{profile}$) at 77 K.

Filament-to-filament resistance measurement refers to forcing current across one single groove at a time, measuring the voltage drop across it. Consequently, $R_{FTF}$ represents the resistance of one single laser striated groove.

In a profile resistance measurement current is forced across all 10 filaments while the voltage drop is measured cumulatively, with reference to the first filament. In this measurement a profile plot is obtained that shows how the resistance of the tape changes along its width and consequently serves with information on where the transversal currents possibly flow.

The resistances were calculated from I-V curve measurements where the voltage was measured by spring pins, using the four-point probe method, placed on the filaments and the current was calculated from a shunt resistance inserted in the circuit. 20 to 40 data points were recorded for Ag-cap and Cu-cap tapes, respectively, using a range of -200 mA to +200 mA DC current and a nanovoltmeter. Hence the values shown in this paper represent statistical averages of 20 to 40 readings.

To explain the transversal coupling current path the previously developed chain resistor model was used, shown in [22]. This model takes into consideration the resistivity of redeposited material after the laser scribing process. In the model $R_b$, $R_{HS,gap}$, $R_{HS,fil}$ represent the barrier resistance, the resistance of the Hastelloy below a laser groove (gap) and the resistance of the Hastelloy below a filament, respectively. Since $R_{HS,gap}$ and $R_{HS,fil}$ only depend on the resistivity and thickness of the substrate (defined by the tape architecture), the only variable is the barrier resistance, $R_b$, which is therefore the deciding factor in the current splitting, shown in Fig. 1.

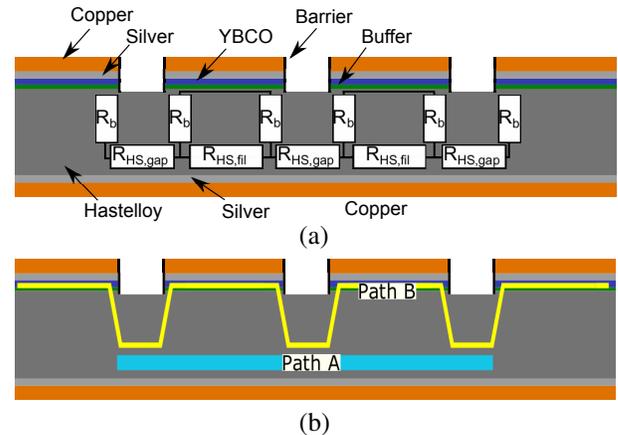

Fig. 1 a) Equivalent circuit of a 4 filament HTS tape (cross-section). b) Possible transverse current paths shown in a 4 filament HTS tape. The image is not to scale.

The filament to filament resistance depends strongly on the scribing process and the tape architecture and therefore it is difficult to give a universal value. Rewriting the formula for transverse resistance for this architecture results in,

$$R = \rho \frac{L}{A} = \rho \frac{w_{groove}}{h_{groove} \times L_{sample}} \qquad (3)$$

Where $\rho$ is the resistivity and $w_{groove}$, $h_{groove}$ and $L_{sample}$ are the groove width, groove height (or depth) and sample length, respectively. The groove height, width and sample length depend on the defined architecture. However, the resistivity also depends strongly on the laser parameters: power, pulse duration, repetitions; and therefore it may not be identical for two tapes with identical architecture, but different laser scribing settings.

## IV. Results

### A. Ag-cap Tapes

In this section the resistance between filaments of non-stabilized, Ag-cap tapes is discussed, with and without oxygenation process as shown in Table I.

#### TABLE I
Average filament to filament (groove) resistance of 15 cm long Ag-cap tapes.

| Company | Surface Finish | $R_{FTF}$ (mΩ) |
|---|---|---|
| SuperPower | - | 4.1 |
| SuperPower | Oxidized | 27.1 |
| SuperOx | - | 7.0 |
| SuperOx | Oxidized | 292.0 |

Table I shows that the filament-to-filament resistance of Ag-cap tapes is in the order of 4 mΩ to 7 mΩ before oxidation and between 30 mΩ and 300 mΩ after oxidation. Fig. 2 shows the individual values of the filament-to-filament resistance measurements and it is clear that in the case of oxidized tapes there can be rather large differences. This can be explained by the quality of the laser scribing as well as the oxidation process.

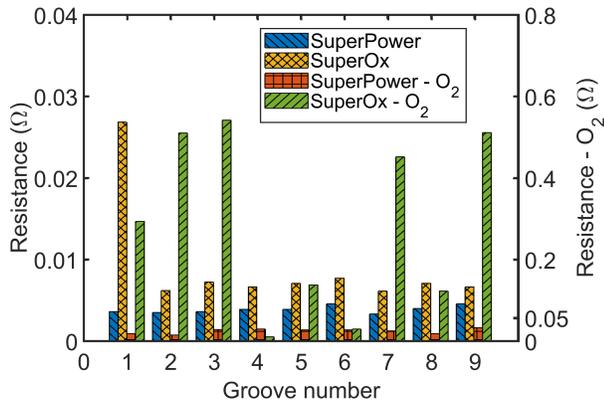

Fig. 2 Individual filament-to-filament resistances of the four, 15 cm long Ag-cap tape samples shown in Table I. $O_2$ represents the oxidized samples.

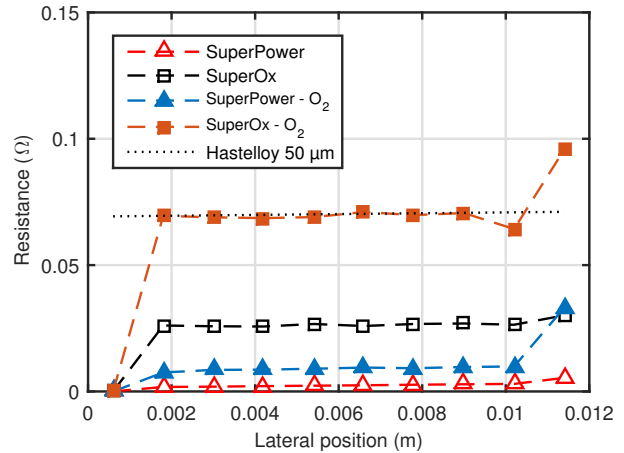

Fig. 3 Profile resistance measurement of 4 samples from SuperOx and SuperPower as previously introduced in Section III. $O_2$ indicates the oxidized samples. For reference the resistance of a 50 µm thick Hastelloy tape is also shown, with an offset to the oxidized SuperOx tape's first measurement point.

Referring to Fig. 1, the measurements equal to barrier resistances in the mΩ range as well according to

$$R_b = \frac{R_{FTF} - R_{HS,gap}}{2} \qquad (4)$$

$R_{HS,fil}$ in Fig. 1 is approximately 195 µΩ, based on a Hastelloy resistivity of $\rho_{HS} = 1.24$ µΩ.m [15] and the dimensions of the samples. The barrier resistance is then at least one order of magnitude larger; hence transverse currents will favour flowing across the lower resistance path A, through the Hastelloy.

This behaviour is shown in Fig. 3 on the profile resistance measurement of the oxidized and non-oxidized Ag samples. In every case an initial and final "step" is observable where the current has to pass through an initial barrier resistance from the top surface of the filament into the body of the tape. As visible, after this the resistance increases rather slowly, which indicates that current flows predominantly along a low resistance path. From the results it is known that this resistance gradient is practically identical to that of the Hastelloy substrate (see offset Hastelloy line in Fig. 3). Then, at the last filament, the current has to close the circuit and overcome another barrier resistance between substrate and silver coating. This results in a final "step" in the resistance profile.

Hence the slope of the middle points, $\Delta R/\Delta x$, of such a plot can be considered as an indication of where the current flows. A slope close to that of the Hastelloy indicates that most of the current flows in the Hastelloy, which causes low coupling. A shallower slope, however, indicates that there is a path less resistive than that of the substrate. This is only possible through a parallel connection of resistors, resulting in current splitting between paths A and B. This then indicates a certain degree of coupling for which higher ac losses are expected.

It is important to mention that the transverse resistivity of such a tape cannot be described with (2), since it assumes a simple series connection of the barriers. Even though this is true in the case where $R_b \cong R_{HS,fil}$, it does not apply to Ag-cap tapes or to tapes where the barrier resistance is considerably larger than the Hastelloy's resistance under the filaments.

### B. Cu-cap Tapes

In copper stabilized tapes the barrier resistance is in the same order of magnitude as the substrate's resistance under the laser groove, as shown in Table II. Hence the current will split between path A and B (Fig. 1) which results in coupling the filaments.

| Company | Copper (µm) | $R_{FTF}$ (µΩ) |
| --- | --- | --- |
| SuperPower | 20 | 33.0 |
| SuperOx | 5 | 685.0 |
| SuperOx[1] | 5 | 5.7 |
| SuperOx | 10 | 68.2 |
| SuperOx | 20 | 34.7 |

Fig. 4 shows two tapes with 5 µm copper stabilization with different preparation steps. The triangular markers show the results for a tape with 5 µm copper stabilizer striated at KIT-ITEP (SAE). The rectangular markers show an Ag-cap tape that was first striated and then electroplated with 5 µm of copper (SBE). The SAE tape shows the previously observed behaviour and initial onset of resistance where current overcomes an initial barrier resistance. The resistance gradient across the transverse direction, however, is lower than that of the pure Hastelloy substrate, indicating a certain level of coupling already with just 5 µm of copper stabilization.

The SBE tape shows no sudden onset of resistance, but has a steady resistance slope, close to that of a 5 µm thick pure copper plate of the same dimensions. This can be explained by the fact that $R_b$ is much smaller than $R_{HS,fil}$ (Table II) and therefore for the transverse currents it is more favourable to flow across path B. As a consequence, significant coupling losses are expected.

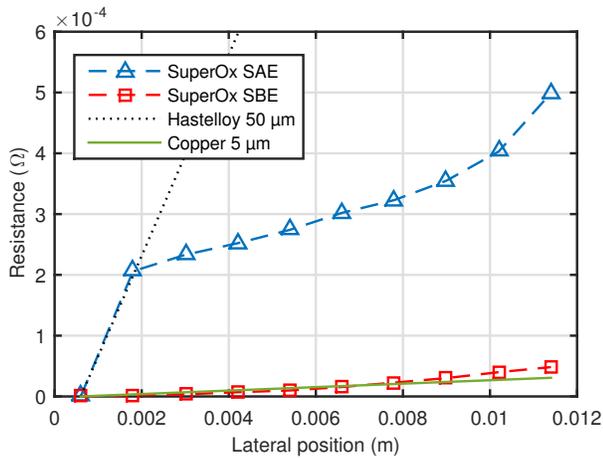

Fig. 4  Profile resistance plot of two SuperOx 5 μm Cu-cap samples. For reference a 50 μm thick Hastelloy and a 5 μm thick copper tape's resistance is also shown.

The performance and degree of filament coupling can be explained by the preparation steps of the two tapes. Fig. 5 and 6 show scanning electron microscopy (SEM) images of the discussed tapes. While the groove of the standard SAE tape is uniform and well-defined, the groove of the SBE tape contains copper islands as well as places fully covered with copper. Such parts – resistive bridges – provide a low resistance path for transverse currents to flow and hence cause the measured lower transverse resistance.

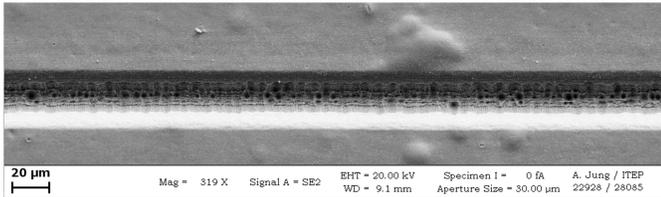

Fig. 5  SEM image of a laser-scribed groove of a 5 μm thick Cu-cap tape striated by cutting through the copper (SAE).

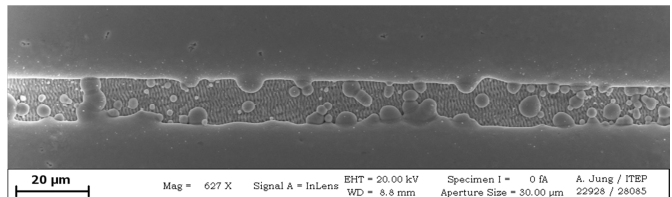

Fig. 6  SEM image of a laser-scribed groove of an Ag-cap tape striated and then electroplated with 5 μm copper (SBE).

### C.  Resistance Profiles of Internal Coupling Current Loops

The effect of externally induced coupling current loops between internal filaments was also investigated for both SuperPower and SuperOx Ag-cap tapes, as illustrated in Fig. 7. This was done by a series of profile resistance measurements where filaments 1-10, 2-9, 3-8 and 4-7 were measured according to Fig. 7.

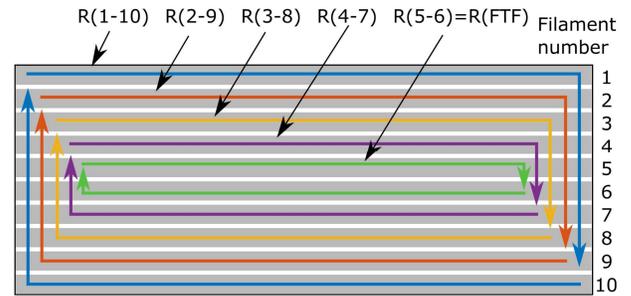

Fig. 7  Schematic illustration of the internal coupling current loops in an HTS tape with 10 filaments. The loops also indicate the most likely paths of transverse currents and hence the profile of transverse resistance, which was measured.

The initial steps caused by current passing through a barrier resistance were observed in both Ag-cap tapes shown in Fig. 8 and Fig. 9. The resistance slope of the intermediary filaments is close to that of the Hastelloy substrate, indicating here as well that path A is preferred for the current. This implies that induced currents would flow in all loops predominantly through the Hastelloy and not across resistive bridges. Additionally, Fig. 9 (and Fig. 2 as well) also shows how the prepared SuperOx tape's first laser groove had an uncharacteristically high resistance. Subsequent grooves have shown a more uniform resistance, indicated by more even initial and final steps.

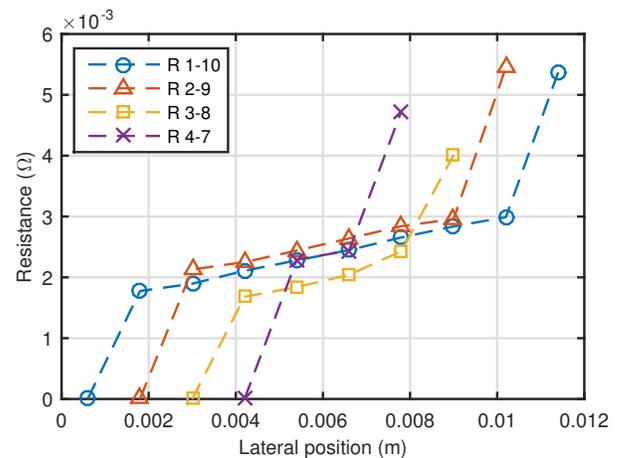

Fig. 8  Resistance profiles of the SuperPower 10 filament Ag-cap tape, corresponding to the probable coupling current loops as defined in Fig. 7. The labels are also identical to that introduced in Fig. 7.

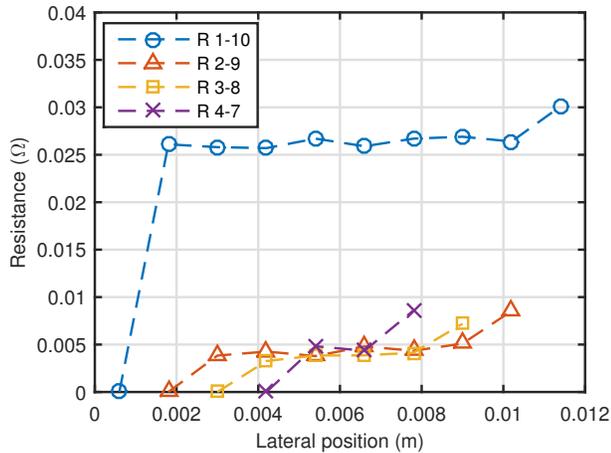

Fig. 9 Resistance profiles of the SuperOx 10 filament Ag-cap tape, corresponding to the probable coupling current loops as defined in Fig. 7. The labels are also identical to that introduced in Fig. 7.

## V. MODELLING

Based on the chain resistor model of transverse resistivity a simple, equivalent circuit equation was derived to describe a tape's behaviour. Taking an $n$ filament HTS tape and starting the circuit from one end of the tape an equation was derived describing any *intermediary* filament. However, the same equation does not describe the last, $n^{th}$ filament's resistance. To calculate the last point's resistance the two edge filaments were taken as fixed points with a repeating structure between them. Therefore the modelled resistance can be written as a function of the filament $N_f$:

$$R_{calc}(N_f) =$$
$$\begin{cases} R_b + \left[ (N_f - 1)\left( R_{HS,gap} + \dfrac{2R_b \times R_{HS,fil}}{2R_b + R_{HS,fil}} \right) \right], & N_f < n \\ 2R_b + \left[ (N_f - 2)\left( R_{HS,gap} + \dfrac{2R_b \times R_{HS,fil}}{2R_b + R_{HS,fil}} \right) \right] + R_{HS,gap}, & N_f = n \end{cases}$$
(5)

where $N_f$ is the actual filament number and $n$ is the total number of filaments.

Using the derived equations the Ag-cap tapes were modelled by using experimentally obtained barrier resistance values. The results are shown in Fig. 10 where the SuperOx tape's resistance was normalized by replacing the uncharacteristically large initial step with an average value. This simple model can approximate reasonably well the behaviour of an *ideal* Ag-cap tape. Despite the occasional differences in the initial and final steps between model and measurements, the resistance slope of the intermediary filaments is accurately calculated. Hence by knowing the average $R_{FTF}$ of a tape, using (4) and (5) the current paths can be approximated.

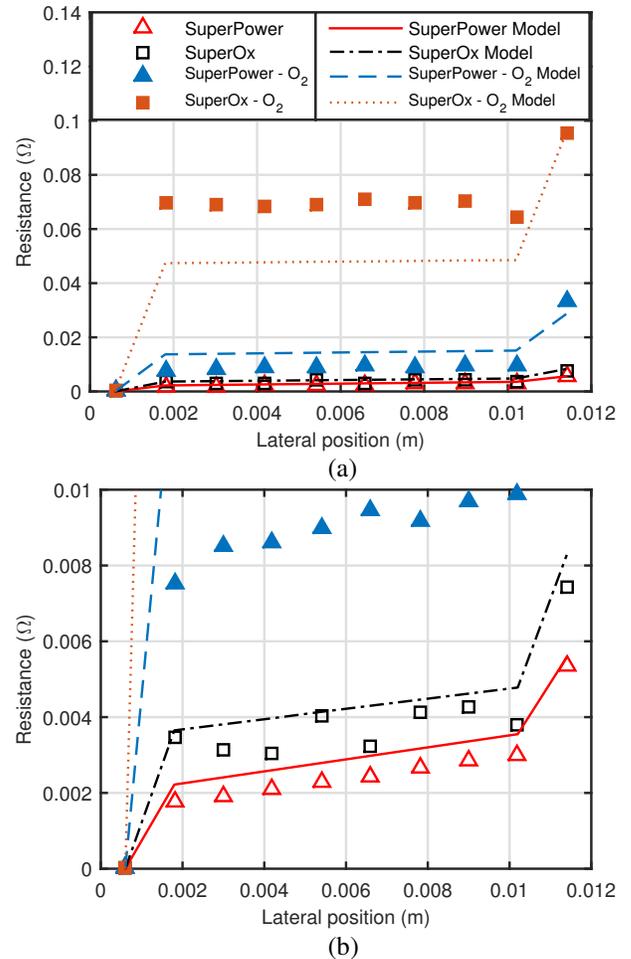

Fig. 10 a) Comparison between model and measured data on four, 15 cm long Ag-cap samples from SuperPower and SuperOx. The $O_2$ index represents the samples that have been oxidized in order to increase the grooves' resistance. b) A zoomed-in section showing a comparison between model and measurement data only for the two, 15 cm long Ag-cap samples (without oxidation).

## VI. CONCLUSION

The transverse resistance of various striated coated conductors with and without stabilization from SuperOx and SuperPower samples was measured. Results indicate that due to the relatively high barrier resistance in Ag-cap samples, in the orders of a few m$\Omega$ up to a few hundred m$\Omega$, transversal currents predominantly flow across the Hastelloy substrate after overcoming the first barrier resistance. In comparison, the barrier resistance in the measured copper stabilized samples was below 1 m$\Omega$, in some occasions under 10 µ$\Omega$. This resistance is low enough to provide an alternative path for the current to flow, which leads across the laser grooves and thereby introduces higher coupling losses in AC applications.

It has also been found that transverse currents in all of the probable coupling current loops in Ag-cap samples would behave similar to each other. In every loop the current has to bypass an initial barrier resistance, however, thereafter the most favourable path is through the substrate layer. This is due to the fact that the encountered resistance along the substrate is lower than that across the grooves (barrier resistance).

Using equivalent circuit equations, derived from the single cell model of transverse resistivity, the profile resistance measurements of SuperOx and SuperPower Ag-cap samples could be modelled with reasonable accuracy. Nevertheless, some discrepancies are present due to uncharacteristic groove resistances, caused most likely by the laser scribing process. The models serve with information about the transversal current paths, which can be derived from $\Delta R/\Delta x$ of the intermediary section of the model. Using this information then one can conclude on the degree of filament coupling in a tape and engineer a certain level of coupling in future samples.

## VII. Acknowledgments

The authors are thankful for the company SuperOx, Moscow, Russia for preparing and analysing special HTS samples used in the measurements.